\documentclass[authorversion,sigconf,nonacm]{acmart}
\AtBeginDocument{%
  \providecommand\BibTeX{{%
    \normalfont B\kern-0.5em{\scshape i\kern-0.25em b}\kern-0.8em\TeX}}}

\acmConference[ACSAC '21]{...}{...}{...}
\acmBooktitle{...}
\acmPrice{...}
\acmISBN{...}

\acmSubmissionID{71}
\usepackage{graphicx}
\graphicspath{{./figures/}}
\usepackage{colortbl}
\usepackage{tikz}
\usepackage{subcaption}
\usepackage{amsmath}
\usepackage{amsmath}
\usepackage{algorithm}
\usepackage[noend]{algpseudocode}
\makeatletter
\def\BState{\State\hskip-\ALG@thistlm}
\makeatother

\usepackage{soul}
\usepackage{makecell}

\usepackage{pifont}%

\usepackage[para,online,flushleft]{threeparttable}
\usepackage{booktabs}
\usepackage{longtable}
\usepackage{pgfplots}
\usepackage{pgfplotstable}
\usetikzlibrary{pgfplots.groupplots}
\usepackage[textsize=scriptsize]{todonotes}

\usepackage{multirow}
\usepackage{listings}
\definecolor{mGreen}{rgb}{.6,0.6,0}
\definecolor{mGray}{rgb}{0.5,0.5,0.5}
\definecolor{mPurple}{rgb}{0.58,0,0.82}
\definecolor{backgroundColour}{rgb}{0.98,0.98,0.99}
\definecolor{shhi}{rgb}{.7,0.7,.7}

\lstdefinestyle{CStyle}{
    backgroundcolor=\color{white},
    commentstyle=\color{mGreen},
    keywordstyle=\color{magenta},
    numberstyle=\tiny\color{black},
    stringstyle=\color{mPurple},
    basicstyle=\footnotesize,
    breakatwhitespace=false,
    breaklines=true,
    captionpos=b,
    keepspaces=true,
    numbers=left,
    numbersep=5pt,
    showspaces=false,
    showstringspaces=false,
    showtabs=false,
    tabsize=2,
    language=C,
    xleftmargin=1em,
    framexleftmargin=1em,
    frame = single,
}

\usepackage[toc,acronym,nowarn]{glossaries}

\newacronym{OS}{OS}{Operating System}
\newacronym{SE}{SE}{Secure Element}
\newacronym{NFC}{NFC}{Near Field Communication}
\newacronym{BLE}{BLE}{Bluetooth Low Energy}
\newacronym{RIL}{RIL}{Radio Interface Layer}
\newacronym{SELinux}{SELinux}{Security-Enhanced Linux}
\newacronym{LSM}{LSM}{Linux Security Modules}
\newacronym{TCB}{TCB}{Trusted Computing Base}
\newacronym{MAC}{MAC}{Mandatory Access Control}
\newacronym{MMIO}{MMIO}{Memory-mapped I/O}
\newacronym{PIO}{PIO}{Programmable I/O}
\newacronym{IPC}{IPC}{Inter-Process Communication}
\newacronym{ICC}{ICC}{Inter-Container Communication}
\newacronym{cgroups}{cgroups}{control groups}
\newacronym{CA}{CA}{Certificate Authority}
\newacronym{PKI}{PKI}{Public Key Infrastructure}
\newacronym{MITM}{MITM}{Man-In-The-Middle}
\newacronym{BYOD}{BYOD}{Bring-Your-Own-Device}
\newacronym{CM}{CM}{Container Management}
\newacronym{SM}{SM}{Security Management}
\newacronym{HAL}{HAL}{Hardware Abstraction Layer}
\newacronym{TLS}{TLS}{Transport Layer Security}
\newacronym{C2C}{C2C}{Container To Container}
\newacronym{Protobuf}{Protobuf}{Protocol Buffers}
\newacronym{SDO}{SDO}{Sensitive Data Object}
\newacronym{VMA}{VMA}{Virtual Memory Area}
\newacronym{PGD}{PGD}{Page Global Directory}
\newacronym{PUD}{PUD}{Page Upper Directory}
\newacronym{PMD}{PMD}{Page Middle Directory}
\newacronym[firstplural=Page Table Entries (PTEs)]{PTE}{PTE}{Page Table Entry}
\newacronym{COW}{COW}{Copy-On-Write}
\newacronym{IV}{IV}{Initialization Vector}
\newacronym{ESSIV}{ESSIV}{Encrypted Salt-Sector Initialization Vector}
\newacronym{KSM}{KSM}{Kernel Samepage Merging}
\newacronym{JIT}{JIT}{Just-In-Time}
\newacronym{DMA}{DMA}{Direct Memory Access}
\newacronym{FDE}{FDE}{Full Disk Encryption}
\newacronym{AS}{AS}{Address Space}
\newacronym{GCM}{GCM}{Google Cloud Messaging}
\newacronym{TPM}{TPM}{Trusted Platform Module}
\newacronym{JTAG}{JTAG}{Joint Test Action Group}
\newacronym{LUKS}{LUKS}{Linux Unified Key Setup}
\newacronym{VPN}{VPN}{Virtual Private Network}
\newacronym{PBKDF2}{PBKDF2}{Password-Based Key Derivation Function 2}
\newacronym{KVM}{KVM}{Kernel-based Virtual Machine}
\newglossaryentry{VM}
{
  name={VM},
  description={Virtual Machine},
  first={\glsentrydesc{VM} (\glsentrytext{VM})},
  plural={VMs},
  descriptionplural={Virtual Machines},
  firstplural={\glsentrydescplural{VM} (\glsentryplural{VM})}
}
\newacronym{HV}{HV}{Hypervisor}
\newacronym{SEV}{SEV}{Secure Encrypted Virtualization}
\newacronym{SME}{SME}{Secure Memory Encryption}
\newacronym{TSME}{TSME}{Transparent \gls{SME}}
\newacronym{SP}{AMD-SP}{Secure Processor}
\newacronym[firstplural=Guest Virtual Adresses (GVAs)]{GVA}{GVA}{Guest Virtual Address}
\newacronym[firstplural=Guest Physical Addresses (GPAs)]{GPA}{GPA}{Guest Physical Address}
\newacronym[firstplural=Host Physical Addresses (HPAs)]{HPA}{HPA}{Host Physical Address}
\newacronym[firstplural=System Physical Addresses (SPAs)]{SPA}{SPA}{System Physical Address}
\newacronym{GPT}{GPT}{Guest Page Table}
\newacronym{HPT}{HPT}{Host Page Table}
\newacronym{TLB}{TLB}{Translation Lookaside Buffer}
\newacronym{PoC}{PoC}{Proof-of-Concept}
\newacronym{ORAM}{ORAM}{Oblivious RAM}
\newacronym{SEV-ES}{SEV-ES}{SEV Encrypted State}
\newacronym{SEV-SNP}{SEV-SNP}{SEV Secure Nested Paging}
\newacronym{RMP}{RMP}{Reverse Map Table}
\newacronym{VMCB}{VMCB}{Virtual Machine Control Block}
\newacronym{SLAT}{SLAT}{Second Level Address Translation}
\newacronym{SSH}{SSH}{Secure Shell}
\newacronym{RSA}{RSA}{Rivest–Shamir–Adleman}
\newacronym{ECDHE}{ECDHE}{Elliptic-Curve Diffie-Hellman Ephemeral}
\newacronym{AES}{AES}{Advanced Encryption Standard}
\newacronym{OOM}{OOM}{Out Of Memory}
\newacronym{MKTME}{MKTME}{Multi-Key Total Memory Encryption}
\newacronym{VMI}{VMI}{Virtual Machine Introspection}
\newacronym{MAD}{MAD}{Median Absolute Deviation}
\newacronym{HSM}{HSM}{Hardware Security Module}
\newacronym{AE}{AE}{Automatic Exit}
\newacronym{NAE}{NAE}{Non-Automatic Exit}
\newacronym{AES-NI}{AES-NI}{AES New Instructions}
\newacronym{NIC}{NIC}{Network Interface Card}
\newacronym{NMI}{NMI}{Non-Maskable Interrupt}
\newacronym{MTU}{MTU}{Maximum Transmission Unit}
\newacronym{VA}{VA}{Virtual Address}
\newacronym{GFN}{GFN}{Guest Frame Number}
\newacronym{SFN}{SFN}{System Frame Number}
\newacronym{IOMMU}{IOMMU}{I/O Memory Management Unit}
\newacronym{AISE}{AISE}{Address Independent Seed Encryption}
\newacronym{MT}{MT}{Merkle Tree}
\newacronym{BMT}{BMT}{Bonsai Merkle Tree}
\newacronym{LPID}{LPID}{Located Page IDentifier}
\newacronym{CB}{CB}{Counter Block}
\newacronym{PRD}{PRD}{Page Root Directory}
\newacronym{SWIOTLB}{SWIOTLB}{Software I/O Translation Buffer}
\newacronym{ASID}{ASID}{Address Space Identifier}
\newacronym{vCPU}{vCPU}{virtual CPU}
\newacronym{VC}{\texttt{\#VC}}{VMM Communication Exception}
\newacronym{GHCB}{GHCB}{Guest Hypervisor Communication Block}
\newacronym{IDT}{IDT}{Interrupt Descriptor Table}
\newacronym{KASLR}{KASLR}{Kernel Address Space Layout Randomization}
\newacronym{SLES}{SLES}{SUSE Linux Enterprise Server}
\newacronym{RHEL}{RHEL}{RedHat Enterprise Linux}
\newacronym{TSC}{TSC}{Time Stamp Counter}
\newacronym{IRET}{IRET}{Return from interrupt}
\newacronym{ROP}{ROP}{Return-oriented programming}
\newacronym{MSR}{MSR}{Model Specific Register}
\newacronym{RNG}{RNG}{Random Number Generator}
\newacronym{PRNG}{PRNG}{Pseudo Random Number Generator}
\newacronym{IBS}{IBS}{Instruction Based Sampling}
\newacronym{VMSA}{VMSA}{\gls{VM} Save Area}
\newacronym{OVMF}{OVMF}{Open Virtual Machine Firmware}
\newacronym{RW}{RW}{Read-Write}
\newacronym{RWX}{RWX}{Read-Write-Execute}
\newacronym{TDX}{TDX}{Trust Domain Extensions}
\newacronym{PCI}{PCI}{Peripheral Component Interconnect}
\newacronym{USB}{USB}{Universal Serial Bus}
\newacronym{ACPI}{ACPI}{Advanced Configuration and Power Interface}
\newacronym{DEP}{DEP}{Data Execution Prevention}
\newacronym[firstplural=Extended Page Tables (EPTs)]{EPT}{EPT}{Extended Page Table}
\newacronym{ShEPT}{Shared EPT}{Shared Extended Page Table}
\newacronym{SeEPT}{Secure EPT}{Secure Extended Page Table}
\newacronym{VE}{\#VE}{Virtualization Exception}
\newacronym{TDMR}{TDMR}{Trust Domain Memory Region}
\newacronym{SEAM}{SEAM}{Secure-Arbitration Mode}
\newacronym{VIRTIO}{VirtIO}{Virtual I/O}
\newacronym{AST}{AST}{Abstract Syntax Tree}
\newacronym{LKL}{LKL}{Linux Kernel Library}
\newacronym{KASAN}{KASAN}{Kernel Address Sanitizer}
\newacronym{ASAN}{ASAN}{Address Sanitizer}
\newacronym{IDE}{IDE}{Integrated Drive Electronics}
\newacronym{ISA}{ISA}{Industry Standard Architecture}

\makeglossaries
\glsdisablehyper

\newcommand{\coolnameplain}{VIA}
\newcommand{\coolname}{{\coolnameplain }}
\newcommand{\kernelversion}{{5.10.0-rc6}}
\newcommand{\kernelversionbugs}{{5.10.8}}
\newcommand{\drvanalyzed}{{22 }}
\newcommand{\drvwithbugs}{{20 }}
\newcommand{\drvanalyzedqvd}{{19 }}
\newcommand{\drvanalyzedcvm}{{3 }}

\newcommand{\nbugs}{{50 }}

\newcommand{\nbugsconf}{{9 }}

\newcommand{\dynamicperf}{{570 }}

\usepackage{tikz}
\newcommand*\circled[1]{\tikz[baseline=(char.base)]{
            \node[shape=circle,draw,inner sep=0.9pt] (char) {#1};}}
\newcommand*\circledg[1]{\tikz[baseline=(char.base)]{
            \node[shape=circle,draw,inner sep=0.9pt,fill=lightgray] (char) {#1};}}
\newcounter{fhcnt}

\begin{document}

\date{}

\title{\coolnameplain: Analyzing Device Interfaces of Protected Virtual Machines}

\author{Felicitas Hetzelt}
\affiliation{%
  \institution{TU Berlin}
  \city{}
  \country{}
  }
\email{file@sect.tu-berlin.de}

\author{Martin Radev}
\affiliation{%
  \institution{Fraunhofer AISEC}
  \city{}
  \country{}
  }
\email{martin.b.radev@gmail.com}

\author{Robert Buhren}
\affiliation{%
  \institution{TU Berlin}
  \city{}
  \country{}
  }
\email{robert.buhren@sect.tu-berlin.de}

\author{Mathias Morbitzer}
\affiliation{%
  \institution{Fraunhofer AISEC}
  \city{}
  \country{}
  }
\email{mathias.morbitzer@aisec.fraunhofer.de}

\author{Jean-Pierre Seifert}
\affiliation{%
  \institution{TU Berlin}
  \city{}
  \country{}
  }
\email{jpseifert@sect.tu-berlin.de}

\begin{abstract}

Both AMD and Intel have presented technologies for confidential computing in cloud environments.
  The proposed solutions --- AMD SEV (-ES, -SNP) and Intel TDX --- protect \glspl{VM} against attacks from higher privileged layers through memory encryption and integrity protection.
  This model of computation draws a new trust boundary between virtual devices and the \gls{VM},
  which in so far lacks thorough examination.
  In this paper, we therefore present an analysis of the virtual device interface
  and discuss several attack vectors against a protected \gls{VM}.
  Further, we develop and evaluate \coolname{}, an automated analysis tool to detect cases of improper sanitization of input recieved via the virtual device interface.
  \coolname{} improves upon existing approaches for the automated analysis of device interfaces in the following aspects:
  (i) support for virtualization relevant buses,
  (ii) efficient \gls{DMA} support and
  (iii)  performance.
  \coolname{} builds upon the Linux Kernel Library and clang's libfuzzer to fuzz the communication between the driver and the device via MMIO, PIO, and DMA.
An evaluation of \coolname{} shows that it performs \dynamicperf executions per second on average
and improves performance compared to existing approaches by an average factor of $2706$.
Using \coolname, we analyzed \drvanalyzed drivers in Linux \kernelversion, thereby uncovering \nbugs bugs and %
initiating multiple patches to the virtual device driver interface of Linux.
To prove our findings' criticality under the threat model of AMD SEV and Intel TDX, we showcase three exemplary attacks based on the bugs found.
The attacks enable a malicious hypervisor to corrupt the memory and gain code execution in protected \glspl{VM} with SEV-ES and are theoretically applicable to SEV-SNP and TDX.

\end{abstract}

\ccsdesc[500]{Security and privacy~Virtualization and security}

\maketitle

\section{Introduction}
Cloud computing provides many advantages, such as on-demand resource allocation and high-availability guarantees.
Nevertheless, many enterprises have not moved their data into the cloud due to security and privacy concerns~\cite{amigorena2019why};
one major concern being the compromise of the \gls{HV} which has full access to the potentially sensitive data processed within the \gls{VM}.

To protect data within cloud environments from a potentially malicious \gls{HV}, several solutions have been proposed by academia~\cite{cloudvisor2011, hsvm2011, hypercoffer2013, hyperwall2012}.
In 2016, AMD proposed the first commercially available solution called \gls{SEV}~\cite{sme2016}, followed later by similar solutions from Intel~\cite{tdx-spec} and IBM~\cite{ibm2018supporting}.
These solutions enhance the security of \glspl{VM}, as the \gls{VM} no longer depends on the \gls{HV}'s security or on the integrity of the cloud provider.
While recent research has uncovered security issues specific to the design of the \gls{SEV} technologies~\cite{cohen2019sev, buhren2019insecure, hetzelt2017security, du2017secure, morbitzer2018severed, morbitzer2019extracting, werner2019severest, li2019exploiting, wilke2020sevurity, li2020crossline}, little research has been done in analyzing the \gls{VM}'s software under this new threat model~\cite{radev2020exploiting}.
Specifically, only limited effort has been spent on analyzing security risks due to drivers which communicate with devices controlled by the now untrusted \gls{HV}.
Drivers inside the \gls{VM} should now treat input from virtual devices as malicious and should implement proper input validation.
Compared to attacks from external hardware devices against the host OS~\cite{Markettos2019, nohl2014badusb},
validation is even more critical
in a protected \gls{VM} as a malicious \gls{HV} has fine-grained control over the \gls{VM}, greatly enhancing the adversary's abilities.
For example, a \gls{HV} is able to intercept the \gls{VM}'s execution at an arbitrary point in time and
can monitor various feedback channels %
to infer the execution state of the \gls{VM}~\cite{li2019exploiting}.

Identification of drivers performing insufficient validation of \gls{HV} controlled input requires the analysis of a vast codebase.
To be able to quickly identify missing input validation in such a vast code base, we present \coolname{}.
\coolname{} is a fuzzing framework targeted to detect missing input validation in device drivers of \glspl{VM}.
Although dynamic analysis tools exist for the device interface,
they require specialized hardware~\cite{Markettos2019} or devices implementations~\cite{Song2019} and target only a limited subset of driver functionality.
Existing device independent approaches target the USB interface~\cite{github2019syzkaller, Peng2020} and  a limited subset of the \gls{PCI}~\cite{agamotto} interface.
Notably, none of the device independent approaches targets the complex \gls{DMA} interface of device drivers,
which has been shown to be a rich source of vulnerabilities in the past~\cite{Markettos2019}.
\coolname{} supports \gls{VIRTIO}, Platform and \gls{PCI} device interfaces --including \gls{DMA}-- allowing \coolname{} to target a large extent of the virtual device driver interface.

In addition to limited bus support, previous approaches require a full \gls{VM} setup and suffer from poor performance due to delays embedded in device drivers,
frequent context switches between the host and the \gls{VM} and
inefficient interrupt injection.
To address these issues, \coolname{} moves the fuzzing setup and the Linux kernel into a userspace program and thereby eliminates the need for a full \gls{VM} setup including userspace and virtualization software.
\coolname{} further increases analysis throughput by applying optimizations to the Linux kernel environment that reduce redundant delays in driver code.
To inject fuzzing data into the virtual device interface, \coolname{} employs \textit{in-process} IO-interception and avoids costly \texttt{VMEXIT}- or pagefault-based approaches.
Further, \coolname{} incorporates information about the target driver state to efficiently guide the analysis through targeted interrupt injection.

Using  \coolname{}, we analyzed \drvanalyzed device drivers.
We identified \nbugs cases of missing input validation
and initiated multiple patches to the virtual device driver interface~\cite{vmxnet3,virtionet0,swiotlb0,virtioring1,virtioring2,virtioring0,virtioring3,virtioring4,gve0,gve1}.
In total, we found issues in \drvwithbugs of the \drvanalyzed analyzed drivers.
To underline the severity of the missing validations, we detail three cases of how a malicious \gls{HV} is able to execute arbitrary code or corrupt protected memory within a \gls{VM}.
To evaluate \coolname{}, we %
compare our approach against Agamotto~\cite{agamotto}, a state-of-the-art, \gls{VM}-based, \gls{PCI} device fuzzer.
Our evaluation shows that, \coolname{} on average improves the analysis throughput by factor of $2706$
and the percentage of driver code covered by a factor of $2.26$.

In summary, our contributions are:
\begin{itemize}
  \item{\textbf{Analysis of the virtual device attack vector.}
    We identify and describe the virtual device interface as a new attack vector against protected \glspl{VM}.
      This new attack vector applies to all technologies that aim to isolate complete commodity operating systems.
      }

    \item{\textbf{Categorization of bug classes.}
      We identify and describe numerous bugs in the virtual device driver interface,
      which we classify by type and severity.
      To demonstrate the severity of the identified bugs,
      we describe three proof of concept exploits to obtain code execution and corrupt memory in a protected \gls{VM}.
      }
    \item{\textbf{Targeted dynamic analysis tool.}
      We design, implement and evaluate \coolname{}, a dynamic analysis tool designed to detect missing sanitization of \gls{HV}-controlled data.
      \coolname{} comprises of a userspace fuzzer with improved bus support and performance.
      We detail three optimizations that were applied to improve dynamic analysis:
      in-process IO interception, delay reduction and targeted interrupt injection.
      }
\end{itemize}

\section{Background\label{sec:background}}

The industry has proposed different security technologies that aim to protect \glspl{VM} in an untrusted environment.
This work focuses on technologies for the x86 architecture: AMD's \gls{SEV} and Intel's \gls{TDX}.
This section provides background on both technologies and discusses the use of devices in virtualized environments.

\subsection{Protected Virtual Machines}
AMD \gls{SEV} is the first commercially available technology aimed at protecting \glspl{VM} from a potentially malicious \gls{HV}~\cite{sme2016}.
Protection is achieved
by encrypting the \gls{VM}'s memory with a secret key, that is unique to the \gls{VM} and inaccessible by the cloud provider.
The memory encryption, key management and \gls{VM}-entry and -exit are handled by a separate highly privileged component:
the \textit{Platform Security Processor (PSP)}. %
Each guest controls the encryption of its memory via a dedicated bit in its private page table.
Unencrypted, also named \textit{shared}, pages can be used to exchange data with the \gls{HV}.
Later, AMD released SEV-ES which enhances \gls{VM} protection by encrypting and integrity-protecting the \gls{VM}'s registers.
Another improvement --- SEV-SNP --- was announced in 2020 which protects the emulation of special instructions, as well as the nested page tables against manipulation.
Intel recently proposed a similar technology to SEV-SNP, called \gls{TDX}~\cite{tdx-spec}.
Notably, all these technologies improve security for the protected \gls{VM} through architectural changes, but do not directly address vulnerabilities in the \gls{VM}'s software.

At the time of writing, only \gls{SEV} and SEV-ES are officially supported in Linux and by available CPUs.
AMD SEV-SNP and Intel \gls{TDX} are not officially supported on currently available CPUs, but patches to open-source projects have been made public for both~\cite{github2020tdx, github2021snp}.

\subsection{Device Virtualization\label{subsec:background_device_virt}}
While the hardware virtualization extensions provide CPU and memory virtualization, device virtualization is performed by the \gls{HV} and requires cooperation from the \gls{VM}.
The \gls{VM} discovers attached devices by traversing known bus addresses and then reading the \textit{identifiers and configuration} advertised by the \gls{HV}.
The \gls{VM} then loads and initializes the corresponding device driver.
The driver proceeds with programming the device through either regular access to \gls{MMIO} or special \gls{PIO} instructions.

While \gls{MMIO} and \gls{PIO} are used to send commands and receive configuration information from the device, the communication of large messages, like network packets, happens through \gls{DMA}.
A driver is responsible for allocating physical memory buffers and providing their physical addresses to the device.
When the device decides to provide a payload to the \gls{VM}, the device copies the data to the \gls{DMA} buffer and notifies the driver that new data is available.
The driver can then read the data and process it.

The transparent use of virtual devices with \gls{TDX} or \gls{SEV} requires changes in guest operating systems such as Linux to accommodate for the fact that the \gls{HV} cannot read a guest VM's private memory.
At the kernel initialization stage, the \gls{VM}'s Linux kernel makes adjustments to the \gls{DMA} and \gls{MMIO} API implementation to use decrypted memory for communication.
The device and the corresponding driver can use the memory for sharing frequently accessed communication structures like buffer descriptors and ring queues without the need
for memory synchronization.
Next, the \texttt{swiotlb} component is enabled, which allocates and manages a large contiguous decrypted memory area.
When a driver uses the streaming \gls{DMA} API to map a buffer without coherency requirements (e.g., \texttt{dma\_map\_single}), the implementation internally returns the physical address of a buffer in the \texttt{swiotlb} decrypted memory region.
When the driver synchronizes the memory using the \gls{DMA} API (e.g., \texttt{dma\_unmap\_single}), the \texttt{swiotlb} implementation copies (\textit{bounces}) the memory from the decrypted memory region to the corresponding private buffer of the driver.

\section{Threat Model}\label{sec:threat}

Our threat model assumes a malicious actor who controls the \gls{HV} and seeks to compromise a protected \gls{VM}.
The malicious actor may have gained access to the \gls{HV} by exploiting a system vulnerability or may be a rogue server administrator.
The \gls{HV} cannot directly read or modify the \gls{VM}'s state, as the \gls{VM}'s memory and register contents are encrypted.
The \gls{VM} runs an up-to-date operating system with a Linux kernel.

The \gls{HV} manages access to hardware devices for the \gls{VM} by providing the appropriate virtualized device in the cloud setting.
The virtual device can be based on \gls{VIRTIO} or on a custom interface, but the \gls{VM}'s Linux kernel must support the device.
The \gls{HV} can configure arbitrary devices during the \gls{VM}'s boot and attach or detach any hot-pluggable device during the lifetime of the \gls{VM}.

Being in control of the virtual device, the malicious \gls{HV} is able to modify any data structure shared with the device.
For example, such structures include descriptor buffers and ring queues commonly used in many devices.
The \gls{HV} can intercept \gls{MMIO} and \gls{PIO} operations and provide any value for \gls{MMIO} reads and the \texttt{IN} instruction.
Additionally, the \gls{HV} is able to intercept and modify any communication between the real hardware device and the \gls{VM}'s driver.
For example, the \gls{HV} can modify the contents of network packets sent to the \gls{VM} via the \gls{DMA} interface.
Further, the \gls{HV} is able to inject interrupts into the \gls{VM}, as necessary for sending a notification from a virtual device to the \gls{VM}'s kernel.

This proposed threat model precisely captures the threat models of AMD \gls{SEV-SNP}~\cite{SNP_whitepaper} and Intel \gls{TDX}~\cite{TDX_whitepaper}.
Any attacks on the proposed threat model are also applicable to both of these technologies.

\section{Design}
\label{sec:design}
In this Section we detail the high level design of \coolname{}.
First, we give an overview of the architecture of \coolname{} as well as \coolname's central components.
Further, we present considerations about how the design avoids false positives during bug discovery and
describe options to extend \coolname{}.

\coolname{} is the first comprehensive approach to detect missing sanitization of data provided by a virtual device to a Linux device driver.
It extends the scope of current hardware interface analysis by targeting \gls{VIRTIO}, \gls{PCI}, and Platform device interfaces.
In addition to opening up these new interfaces for analysis, \coolname{} also addresses performance issues in existing approaches.

The dynamic analysis framework %
comprises of a feedback-driven, in-process, userspace fuzzer that mutates data sent from a virtual device to the device driver.

On a high level, \coolname{} provides \textit{device simulations} of varying complexity %
that forward random data to the \gls{HV} interface of device drivers.
The drivers are loaded as shared libraries into a userspace program that provides a Linux kernel environment.
\coolname{} extends and modifies the Linux kernel environment to support the analysis.
Namely, \coolname{} applies \textit{delay reduction}, \textit{targeted interrupt injection} and \textit{in-process} IO-interception to improve fuzzing throughput significantly.
Further, \coolname{} extends the fuzzer's \textit{input generation} mechanism to support the analysis and avoid deadlocks in the driver code.

\begin{figure}
  \centering
  \def\svgwidth{0.8\columnwidth}
  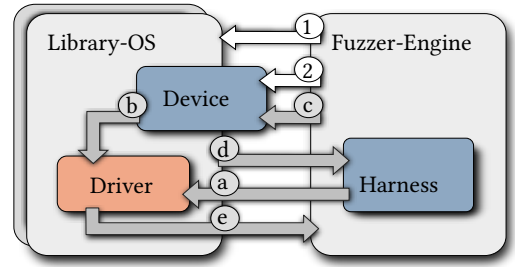
  \caption{
  High-level overview of \coolname{}.
  White arrows indicate initialization, gray arrows indicate actions performed during each fuzzing iteration.
  The main components are the Fuzzer-Engine and the Library-OS, which again contain
  the Harness and the analyzed Driver that is bound to the according virtual Device.
  }
  \label{fig:designdynamic}
\end{figure}

\paragraph{System Overview}
Figure \ref{fig:designdynamic} illustrates the overall design of \coolname{}'s dynamic analysis framework.
The two main components are the \textit{Fuzzer-Engine} and the \textit{Library-OS}.
The Fuzzer-Engine is responsible for setup, monitoring, and controlling of the test execution.
It processes code coverage feedback and provides mutated data that the analyzed \textit{Driver} consumes.
Additionally, it contains an exchangeable \textit{Harness} that drives interaction between the simulated \textit{Device} and the driver.
The Library-OS component contains the Linux kernel code within which we execute the driver,
and software implementations for \gls{PCI}, \gls{VIRTIO}, and Platform devices
that provide mutated data to the driver.

In order to start the analysis, the Fuzzer-Engine first starts one or more instances of the Library-OS \circled{1} and
configures the software device of each instance according to the selected device type \circled{2}.

During each fuzzing iteration, the Fuzzer-Engine executes the instructions provided by the harness.
The harness interacts with the Library-OS via the standard Linux system call interface \circledg{a},
which we extended to facilitate interrupt generation and device driver (un-)initialization.
The role of the harness is to issue a set of system calls to
(i) initialize the driver,
(ii) trigger interactions between the driver and the device \circledg{b} and
(iii) un-initialize the driver.
When the driver requests data from the device,
the device will either forward mutated data from the Fuzzer-Engine \circledg{c} or handle the request according to its internal device simulation logic.
In order to resolve locks quickly,
the Library-OS notifies the harness if a thread initiated by the driver is waiting on a resource \circledg{d};
the harness then has the option to trigger an interrupt in order to resolve the block.
After each execution of the harness, the code coverage generated by the driver is forwarded to the Fuzzer-Engine \circledg{e} to guide future mutations.

\paragraph{Device Simulation}
\coolname{} simulates virtual devices in order to communicate with the \gls{HV} interface of device drivers.
The Fuzzer-Engine configures the device according to a configuration file provided by the analyst.
Depending on the device type (\gls{PCI}, \gls{VIRTIO}, or Platform), \coolname{} uses different approaches to simulate the device.
The device simulation generally falls into one of two categories: \textit{passthrough} and \textit{emulation}.

Both passthrough and emulated devices perform an initial device \textit{specialization} based on a configuration file.
The initial specialization configures device identifiers that allow matching the driver to the device.
In the case of Platform devices, drivers are matched via their name~\cite{platformdevs};
\gls{PCI} and \gls{VIRTIO} devices are matched via Vendor- and Device-IDs~\cite{pcispecconf,virtiospec}.
Additionally, \coolname{} configures the device's resources at this stage, namely interrupts and \gls{MMIO} and \gls{PIO} regions that the driver expects to be provided by the device.

\coolname{} uses passthrough for \gls{PCI} and Platform devices. %
After the initial configuration, the simulated device only forwards mutated data to the driver and
ignores data sent from the driver to the device.
Interrupts have to be triggered by the harness.

The emulation mode allows \coolname{} to target higher layers of the driver code.
To that end, emulated devices adhere to the interface specification (e.g., ~\cite{virtiospec})
to establish basic communication channels between driver and device.
Therefore, emulated devices process data sent from the driver to the device and automatically trigger interrupts depending on the device's internal state.
\coolname{} can simulate \gls{VIRTIO} devices in both passthrough and emulation mode.
While passthrough simulation targets the lower levels of the \gls{VIRTIO} driver stack, e.g., \texttt{virtio-ring} or \texttt{virtio-mmio},
emulation can reach code in higher layers, including drivers such as \texttt{virtio-net} and \texttt{virtio-blk}.
\coolname{} provides a single unified emulated \gls{VIRTIO} device stub to analyze arbitrary \gls{VIRTIO} drivers.

\paragraph{Targeted Interrupt Injection}
Most driver-device interaction is initiated via the execution of interrupt handlers that are part of the driver.
The harness therefore injects interrupts to trigger the interaction between the simulated device and the driver.
However, interrupts should not be injected randomly as this might decrease performance as well as coverage stability.
\coolname{} addresses this issue by providing callbacks to inject interrupts precisely when an execution thread of the driver enters a blocked state
while waiting on data from the device.

\paragraph{Delay Reduction}
Driver execution is often prolonged during
initialization or peripheral input processing.
This is mostly due to properties of the real hardware device that the driver assumes to interact with.
Through manual analysis of driver code, we found these delays to be mostly redundant for software-based device implementations.
Therefore, \coolname{} modifies common kernel interface functions to
eliminate delays in a generic way without impacting the functionality of the driver.
Note that during such delays the processor is not necessarily idle, as the Linux kernel might schedule other workloads intermediately.
Therefore, approaches such as presented in ReVirt~\cite{Dunlap2002}, that skip cpu idle time, are not directly applicable.

\paragraph{Intercepting IO}
In order to inject data into the hardware interface, we have to intercept the normal communication path between driver and device.
\gls{MMIO} and \gls{PIO} regions are accessed via a defined kernel interface.
\coolname{} extends the functions comprising this interface to inject test data during a fuzzing iteration.

For \gls{DMA}, we have to distinguish between coherent and streaming \gls{DMA}.
Streaming \gls{DMA} is synced once before and after device interaction;
therefore, the whole \gls{DMA} buffer is filled by the fuzzer at CPU synchronization points.
Coherent \gls{DMA} can be accessed continuously without explicit synchronization from the guest CPU.
Therefore, fresh values should be provided at each access.
However, the kernel does not offer an interface to hook such accesses. %
\coolname{} instead facilitates the address sanitizer~\cite{asan} instrumentation to intercept accesses to coherent \gls{DMA} memory. %
Address sanitization adds instrumentation around memory accesses that manages values stored in a shadow memory map.
Upon memory access the values in the shadow memory are modified to indicate the allocation state of the accessed memory area.
Further each access is validated by examining the allocation state based on the values stored in the shadow memory.
\coolname{} marks the shadow memory corresponding to coherent \gls{DMA} memory areas with a special value, which is processed by the address sanitizer error handling code.
When handling read operations, the sanitizer's error handling code is augmented to inject the next value from the input stream at the faulting \gls{DMA} address and then continue execution.

To improve performance, some memory access types are excluded from address sanitization.
These include accesses to allocations on the current stack frame that are determined to be in bounds or promotable to register accesses and accesses to certain types of global variables.
We found that, for the analyzed drivers, none of the consistent \gls{DMA} allocations met these criteria.

Kernel addresses exposed to the \gls{HV} can leak information about the address space layout of the \gls{VM}.
To detect such issues, all values exposed to the virtual device via \gls{MMIO}, \gls{PIO} or \gls{DMA} are checked.
If a value contains a kernel address a warning is issued.

\paragraph{Input Generation}
Each fuzzer-generated input is treated as a serialized sequence of memory transfers from the device to the driver.
Each time the driver requests data from the device,
\coolname{} consumes the corresponding number of bytes from the test input.
If the generated input is exhausted, the input is extended with pseudo-random data.
This ensures that the driver receives different values to avoid infinite loops and deadlocks.

\paragraph{Avoiding false positives}
The adaptions of the Linux kernel environment were carefully designed to avoid false positives during bug discovery.
To that end, the framework mimics the functionality of the \texttt{swiotlb} implementation in Linux, which is always enabled with \gls{SEV} and \gls{TDX}.
Like the Linux \texttt{swiotlb} implementation, \coolname{} keeps track of mapped \gls{DMA} memory areas and checks if these areas are currently mapped before the unmap operation can be performed.
If a \gls{DMA} address has not been mapped previously the unmap invocation is rejected.
Without such validation, drivers that pass device controlled \gls{DMA} addresses to \texttt{dma\_direct\_sync\_single\_for\_cpu} or \texttt{dma\_unmap\_single},
such as \texttt{virtio-ring}, \texttt{sungem} and \texttt{sunhme},
could be exploited to corrupt the \gls{VM}'s memory.
Note that \coolname{} uncovered bugs in the \texttt{swiotlb} implementation, as detailed in Section~\ref{sec:case-study}.
To aid the analysis, \coolname{} handles such issues and reports them.

While delay reduction could potentially lead to the detection of bugs that are not exploitable if the delays are re-introduced,
the \gls{HV} has fine grained control over the \gls{VM}'s execution and timers~\cite{radev2020exploiting}.
Therefore the \gls{HV} can adjust scheduling and delays within the \gls{VM} in order to aid the exploitation of such bugs.
Since the \gls{HV} controls interrupt generation under \gls{SEV} as well as \gls{TDX},
no false positives can be generated due to targeted interrupt injection.

\paragraph{Extending \coolname{}}

\coolname{}'s modular design allows for the automated analysis of a wide range of different device driver types.
While this work focuses on \gls{PCI}, \gls{VIRTIO}, and Platform devices,
\coolname{} is designed to be easily extensible to support other device types.

\begin{table}
  \begin{center}
    \caption{Implementation complexity of harnesses and device simulations provided by \coolname{}.
    All code is written in \texttt{C}. The number of lines of code is obtained using the tool \texttt{cloc}~\cite{cloc}}
    \label{tbl:loc}
    \begin{minipage}{.5\linewidth}
      \label{tbl:loc}
      \centering
      \begin{tabular}{ l r }
       \toprule
         Device Type & LoC \\
       \midrule
         \gls{PCI} & 242 \\
         Platform & 79 \\
         \gls{VIRTIO} & 554 \\
       \bottomrule
      \end{tabular}
    \end{minipage}%
    \begin{minipage}{.5\linewidth}
      \begin{tabular}{ l r }
       \toprule
         Harness & LoC \\
       \midrule
         Default & 20 \\
         Network & 73 \\
         Block/Char & 72 \\
       \bottomrule
      \end{tabular}
    \end{minipage}
  \end{center}
\end{table}

Table~\ref{tbl:loc} presents the lines of code required to implement harnesses and device simulations.
Depending on the internal complexity of the simulated device, the implementation can be very minimal as for \gls{PCI} and Platform devices,
or require higher implementation effort as for \gls{VIRTIO}.
To add a new device driver of an already supported device type, the respective driver has to be enabled in the Linux kernel configuration and a configuration file containing the device identifiers has to be created.
\coolname{} already provides a default harness implementation that performs one initialization and un-initialization of the driver in each fuzzing iteration.
This default harness performs a basic analysis of arbitrary device drivers.
Additionally, \coolname{} provides more elaborate harness implementations for network, block, and character drivers,
that interact with resources created by the driver and cover a larger extent of the driver's code base.

\section{Implementation}
In this section, we present the implementation details for \coolname{}. %
Further, we detail the adaptions to the Linux kernel environment that enable the functionality described in Section~\ref{sec:design}.

\label{ch:fuzz_impl}
\coolname{} is based upon \gls{LKL}~\cite{Purdila2010} and clang's libfuzzer~\cite{libfuzzer}.
The basic fuzzing infrastructure is provided by libfuzzer, which handles
mutation, test-case scheduling, as well as test-execution and feedback.
LKL provides the Linux kernel execution environment to load and interact with device driver code.

\paragraph{Instrumentation}
To collect coverage feedback, we instrument the driver code with the recommended clang sanitizer to be used with libfuzzer: \texttt{-fsanitize=fuzzer-no-link}.
To avoid feedback pollution, we only enable feedback collection on the target driver, omitting the remaining kernel code.
To also enable address sanitization instrumentation on the whole kernel, we use the clang flags \texttt{-fsanitize=address -fsanitize-recover=address}.
The address sanitization provided by clang requires the C standard library (libc) to be used for heap management.
Since \gls{LKL} uses the Linux kernel allocation algorithms instead of the one provided by libc,
we adapted the kernel address sanitizer to directly modify the shadow memory maintained by clang's userspace address sanitizer instead
of maintaining its own separate shadow memory region.

\paragraph{Delay Reduction}
In order to avoid unnecessary delays while fuzzing, the kernel's \texttt{*delay}, \texttt{*sleep}, \texttt{schedule\_timeout[\_*]}, and \texttt{time\_before/after} functions are modified when they are invoked from driver code.
In the case of \texttt{schedule\_timeout}, \texttt{*delay}, and \texttt{*sleep} the framework returns immediately.

In the case of \texttt{time\_before} and \texttt{time\_after}, the framework always returns that the queried time is after the specified time stamp.
Timeouts, used in \texttt{wait\_event\_*} functions, are not disabled because they are necessary for the functionality of the driver.
Additionally, the delay in \texttt{queue\_delayed\_work} is always zeroed-out.

\paragraph{Lock Tracking}

To track blocked threads waiting for events, the functions \texttt{wait\_for\_completion\_*} and \texttt{*\_wait\_event\_*} are modified.
Specifically, the functions are extended to store and remove handles to the blocked thread before and after completion of the event.
Whenever a new entry is added, the harness is notified.

\section{Evaluation}\label{sec:eval}
In this section, we first categorize the bugs uncovered in the virtual device interface
and present three exploits based on the discovered bugs.
Finally, we evaluate the performance of \coolname{}.

\subsection{Bug Overview}
\begin{table}
	\caption{List of examined drivers with their type,
	communication interface and number of discovered bugs.
	\textit{QVD} stands for QEMU Virtual Device,
	and \textit{CVM} for Google Confidential \gls{VM}.
	We distinguish between low and high severity bugs.
	An analysis of each bug can be found in our repository~\cite{fuzzsarepo}.}

\begin{center}
\begin{threeparttable}%
\begin{tabular}{ l c c c c c }
 \toprule
	\multirow{2}{*}{Driver} & \multirow{2}{*}{Usage} & \multirow{2}{*}{Type} & \multirow{2}{*}{Bus} & \multicolumn{2}{c}{\#Bugs} \\
			     &                     &                    &                   & Low & High \\
 \midrule
	virtio\_ring & QVD & Common & VIO & 2 & 4 \\
	virtio\_net & QVD & NET & VIO & 1 & 1 \\
	virtio\_blk & QVD & BLK & VIO & 5 & 1 \\
	virtio\_crypto & QVD & CRYPTO & VIO & 2 & 0 \\
	virtio\_rng & QVD & RNG & VIO & 0 & 1 \\
	virtio\_console & QVD & MISC & VIO & 1 & 0 \\
	virtio\_balloon & QVD & MISC & VIO & 2 & 1 \\
	\rowcolor{lightgray} virtio\_input & QVD & MISC & VIO & 0 & 0 \\
	rocker & QVD & NET & PCI & 1 & 3 \\
	sungem & QVD & NET & PCI & 1 & 0\textsuperscript{*} \\
	sunhme & QVD & NET & PCI & 2 & 0\textsuperscript{*} \\
	8139cp & QVD & NET & PCI & 1 & 0 \\
	vmxnet3 & QVD & NET & PCI & 0 & 5 \\
	\rowcolor{lightgray} ne2k-pci & QVD & NET & PCI & 0 & 0 \\
	e100 & QVD & NET & PCI & 0 & 1 \\
	e1000 & QVD & NET & PCI & 0 & 3 \\
	e1000e & QVD & NET & PCI & 1 & 0 \\
	qemu\_fw\_cfg & QVD & MISC & PLT & 1 & 0 \\
	acpi & QVD & MISC & PLT & 0 & 1 \\
	gve & CVM & NET & PCI & 2 & 3 \\
	nvme & CVM & BLK & PCI & 1 & 1 \\
	tpm\_tis & CVM & TPM & PLT & 0 & 2 \\
 \bottomrule
\end{tabular}
\end{threeparttable}%
\end{center}
  \begin{tablenotes}[flushleft]
    \item \scriptsize{\textsuperscript{*} Device-controlled values passed to \texttt{swiotlb\_tbl\_unmap\_single} as described in
Section 6.2 were not counted as separate bugs.}
  \end{tablenotes}
\label{tbl:all_bugs}
\end{table}
\coolname{} was used to examine a total of \drvanalyzed drivers in the Linux kernel with version \kernelversion{}.
Of the analyzed drivers, \drvanalyzedqvd belong to QEMU virtual devices and \drvanalyzedcvm to specialized hardware available in Google's SEV-enabled confidential computing VM~\cite{googlecvm}.
The analysis was performed via intermittent runs over the course of several months (i.e., from November 2020 to February 2021). %
Each fuzzing session usually lasted between a few minutes to several hours.
We responsibly disclosed found security issues with the respective maintainers.
To prepare the drivers for the dynamic analysis, we patched problematic code where possible by adding proper sanitization.
Additionally, we removed complex checks (e.g., checksum comparisons) on data provided by the fuzzer.
This optimization was applied to the drivers \texttt{e1000}, \texttt{rocker} and \texttt{sungem}.
Addressing such issues in a generic way is an orthogonal problem that has been addressed in other works~\cite{Aschermann2019,Chen2019}.
The required modifications were easy to identify from the kernel or crash log, and workarounds could easily be developed.
We list the changes to the driver code in our public repository~\cite{fuzzsarepo}.

In total, \coolname{} uncovered \nbugs novel bugs across the drivers listed in Table~\ref{tbl:all_bugs}.
At the time of writing, \nbugsconf of the discovered bugs in
\texttt{swiotlb},
\texttt{gve},
\texttt{virtio\_ring},
\texttt{virtio\_net} and
\texttt{vmxnet3}
have been, or are currently being fixed~\cite{vmxnet3,virtionet0,swiotlb0,virtioring1,virtioring2,virtioring0,virtioring3,virtioring4,gve0,gve1}.
The applied patches include simple corrections of
copy paste errors~\cite{virtioring1} or
return codes~\cite{virtionet0},
additions of \texttt{NULL} pointer checks~\cite{gve1} and
corrections of driver state updates~\cite{gve0, virtioring0}.
More elaborate patches were applied to facilitate the exlcusion of control data such as pointers,
indices or variables determining access length from device control~\cite{vmxnet3, swiotlb0, virtioring2, virtioring3, virtioring4}.

Each entry in Table~\ref{tbl:all_bugs} was carefully validated by examining the source code of the offending driver, and bug duplicates were removed.
A thorough analysis of each bug is available in our repository~\cite{fuzzsarepo}.
The types of bugs covered include out-of-bounds access, device-shared pointer, invalid memory access, slab management bugs, miscellaneous bugs, assertion failure, unbound memory allocations, and deadlocks.
We distinguish between bugs of high and low severity.
We consider bugs as severe that potentially leak or corrupt guest memory either on their own or by aiding exploitation of different vulnerabilities.
These include out-of-bounds access, device-shared pointer and slab management bugs, such as use-after-free or double-free.
The remaining bug classes are considered low-severity, as they will impact the guest system's availability but likely do not undermine the security guarantees of \gls{SEV} or \gls{TDX}.
Low-severity bugs usually arise from implementation errors, that should be fixed to guarantee proper functionality of the driver.

\begin{table}
\caption{
  Distribution of bug classes found by \coolname{}.
}
\begin{center}
\begin{tabular}{ l c }
 \toprule
 Bug Class & Count \\
 \midrule
 Out-of-Bounds access & 14 \\
 Invalid memory access & 10 \\
 Slab management & 8 \\
 Device-shared pointer & 5 \\
 Miscellaneous & 3 \\ %
 Assertion failure (\texttt{BUG}) & 4 \\
 Unbounded allocation & 5 \\
 Deadlock & 1 \\
 \bottomrule
\end{tabular}
\end{center}
\label{tbl:fuzz_bugs}
\end{table}

Table~\ref{tbl:fuzz_bugs} lists the bug classes and the corresponding number of discovered bugs.
Among the most common bugs are \textit{Out-of-Bounds access} and \textit{Invalid memory access} bugs, with respectively $14$ and $10$ incidents.
The discovered root cause of these two classes is missing validation for the index and length fields of shared data structures.
Next are \textit{slab management} bugs encompassing $8$ instances of use-after-free or double free issues.
Such bugs often occur in the teardown code of a function if error handling is incorrectly implemented.
Further, \coolname{} uncovered $5$ \textit{device-shared pointer} bugs where a \gls{VM} kernel pointer is stored in shared memory. %
These device-shared pointer issues were all present in few network drivers (\texttt{GVE}, \texttt{e100}, \texttt{rocker}, \texttt{vmxnet3}).
The \textit{miscellaneous} bug category comprises of $3$ unique bugs including missing synchronization of device memory, division by zero and limited directory traversal.
Lastly, our fuzzing setup caught $4$ cases of assertions failures, $5$ unbounded memory allocations and $1$ deadlock.
The \textit{assertion failure} class includes cases when the kernel would detect an unexpected state and call \texttt{BUG()}.
The \textit{unbound memory allocations} category considers cases for which the kernel performs large allocations or fails to free allocated memory, leading to out-of-memory conditions.
\textit{Deadlocks} can occur, if data from the device introduces a \gls{VM} state in which a kernel thread is indefinitely waiting for a resource.

\subsection{A Case Study of Three Bugs}\label{sec:case-study}

In this section we discuss how a malicious administrator, with only access to the QEMU process, can exploit
three of the discovered bugs to overwrite secret memory of a virtual machine and to gain code execution inside the virtual machine.

The exploitation of the three bugs was performed on a virtual machine protected with \gls{SEV-ES} and running Linux \kernelversionbugs. %
The exploits were designed with \gls{KASLR} disabled, though a malicious \gls{HV} can use leaked \gls{VM} kernel pointers to
infer the memory layout of the \gls{VM} if required.

\paragraph{Device-shared pointer in vmxnet3}
\begin{figure}[htbp]
    \begin{lstlisting}[style=CStyle, escapechar=!]
struct vmxnet3_rx_buf_info {
	struct sk_buff *skb; // shared ptr with device
	...
};
struct sk_buff {
	void (*destructor)(struct sk_buff *skb);
	...
};
void skb_release_head_state(struct sk_buff *skb) {
	if (skb->destructor) skb->destructor(skb);
}\end{lstlisting}
	\caption{Vulnerable code in the \texttt{vmxnet3} driver. %
	The pointer \texttt{skb} on Line 2 is accessible by the device and can be overwritten.
	By creating a fake \texttt{skb} object and overwriting the pointer, the device can trick the driver into calling any function on Line 12.}
\label{code:vmxnet3_desc}
\end{figure}

The \texttt{vmxnet3} interface relies on an array of descriptors that can be accessed by the device.
A \texttt{vmxnet3} descriptor, shown in Figure~\ref{code:vmxnet3_desc}, contains an associated \texttt{skb} kernel pointer among other fields.
The pointer points to a pre-allocated \texttt{sk\_buff} object in the \gls{VM}'s private memory, which can be used immediately when a network packet is received.

The exposure of this pointer to an untrusted device leaks information about the VM kernel's randomized address space and additionally allows a malicious \gls{HV} to overwrite the pointer with a convenient value.
When the packet is received, the contents of the \texttt{sk\_buff} are parsed, and the object is eventually freed.
Before freeing the packet, the \texttt{sk\_buff}'s destructor would be called, as shown on Line 12.

 In order to exploit the bug the \gls{HV}
overwrites the \texttt{skb} pointer in the descriptor to point to decrypted memory at the moment of sending a packet to the \gls{VM}.
At the same time, the \gls{HV} forges a fake \texttt{sk\_buff} object at the new address and overwrites the destructor pointer to a suitable code gadget in the \gls{VM}'s kernel.
The code gadget would position the stack into decrypted memory, which would then allow continuing execution using return-oriented programming.
We determined empirically that many code gadgets are suitable destructor pointers because many registers and the stack contain the address of the \texttt{skb} in decrypted memory.
Similar exploitation opportunities were found in the driver implementations of \texttt{GVE}, \texttt{e100}, and \texttt{Rocker}.

\paragraph{Use-after-Free bug in \gls{VIRTIO}-Net}
\begin{figure}[htbp]
  \begin{lstlisting}[style=CStyle]
static int virtnet_probe(struct virtio_device *vdev) {
  ...
  struct net_device *dev;
  struct virtnet_info *vi;
  ...
  dev = alloc_etherdev_mq(sizeof(struct virtnet_info), max_queue_pairs);
  ...
  vi = netdev_priv(dev); // vi lies within net_device memory area
  ...
  if (virtio_has_feature(vdev, VIRTIO_NET_F_MTU)) {
    mtu = virtio_cread16(vdev, offsetof(struct virtio_net_config, mtu));
    if (mtu < dev->min_mtu) { // check fails
      goto free;
  ...
free:
  free_netdev(dev); // struct net_device is freed
  return err;       // err == 0
}
static void virtnet_remove(struct virtio_device *vdev)
{
  struct virtnet_info *vi = vdev->priv;
  ...
  unregister_netdev(vi->dev); // vi->dev is device controlled
  ...
}
\end{lstlisting}
	\caption{
	  The \gls{HV} can induce \texttt{virtnet\_probe} to fail by providing a low \texttt{mtu} value.
	  The driver fails to set \texttt{err} correctly.
	  This leads to access of the freed memory, pointed to by \texttt{vi}, during device removal.}
\label{code:virtio_probe}
\end{figure}

\begin{figure}[htbp]
  \begin{lstlisting}[style=CStyle]
static void e1000_dump_eeprom(struct e1000_adapter *adapter) {
  ...
  eeprom.len = ops->get_eeprom_len(netdev); // HV controls eeprom.len
  ...
  data = kmalloc(eeprom.len, GFP_KERNEL); // data will overlap virtnet_info
  ...
  ops->get_eeprom(netdev, &eeprom, data); // HV controlled data is copied
  ...
  kfree(data);
}
\end{lstlisting}
	\caption{
	  The function \texttt{e1000\_dump\_eeprom} is used to gain control over the memory freed in \texttt{virtnet\_probe}.
	  The \gls{HV} controls \texttt{eeprom.len}.
	  The function \texttt{get\_eeprom} fills the allocated buffer with data controlled by the \gls{HV}.}
\label{code:e1000}
\end{figure}
The \texttt{virtio\_net} driver fails to return an error code if the initialization fails due to an invalid \texttt{mtu} value provided by the \gls{HV}.
This leads to a \textit{use-after-free} bug when the device is removed.
The vulnerable code is depicted in Figure~\ref{code:virtio_probe}.
To exploit this bug the \gls{HV} first attaches the \texttt{virtio\_net} device and causes \texttt{virtnet\_probe} to fail by providing a low \texttt{mtu} value.
To obtain control over the freed memory area, the \gls{HV} then attaches the \texttt{e1000} device
and triggers the code path shown in Figure~\ref{code:e1000}.
Specifying an \texttt{eeprom.len} of $0x1000$ will cause a new allocation in the \textit{kmalloc-4k} cache, which overlaps the memory area of the freed \texttt{struct virtnet\_info}.
The structure \texttt{virtnet\_info} contains the field \texttt{dev}, which points to a \texttt{net\_device} structure.
The \texttt{net\_device} structure contains the function pointer \texttt{ndo\_uninit}, that will be called in \texttt{unregister\_netdev} via the invocation of \texttt{virtnet\_remove}.
Similar to the previous exploit, the \gls{HV} overwrites \texttt{virtnet\_info->dev} to point to a decrypted memory area and sets \texttt{ndo\_uninit} to point to a suitable code gadget.
Next the \gls{HV} detaches the \texttt{virtio\_net} device to trigger the vulnerable code path and gain code execution.

The reliability of the exploit relies on two factors.
First, the allocation of the data allocated in \texttt{e1000\_dump\_eeprom} has to overlap the freed \texttt{struct net\_device}.
Second, the re-allocated memory area should not be overwritten before \texttt{virtnet\_remove} is called.
Under \gls{SEV}-(-ES, -SNP) the \gls{HV} has read access to encrypted memory.
This allows the \gls{HV} to infer the host virtual addresses of both allocations by monitoring changes to encrypted memory at suitable points during the interaction between the device and the driver\footnote{Alternatively, page fault side channels~\cite{li2019exploiting} could be used to infer address information.}.
In order to verify whether the data was overwritten,
the \gls{HV} compares the encrypted memory state after \texttt{e1000\_dump\_eeprom} to the current state before detaching the \texttt{virtio\_net} device.
If the inferred addresses differ or the memory has been overwritten
the \gls{HV} can either repeat the exploit attempt by
attaching a new device, or choose to abort the attempt.

\paragraph{Out-of-Bounds bug in \gls{VIRTIO} / swiotlb}
Every \gls{VIRTIO} driver in Linux relies on a common \gls{VIRTIO} interface implementation that handles the state of each communication queue.
Each queue is associated with an array of descriptors, depicted in Figure~\ref{code:virtio_desc}.
When a device sends a payload to the driver, the device typically only reads but does not update the \texttt{addr} and \texttt{len} fields since they are provided by the driver.

\begin{figure}[htbp]
    \begin{lstlisting}[style=CStyle]
struct vring_desc {
	__virtio64 addr;
	__virtio32 len; // not validated.
	__virtio16 flags;
	__virtio16 next;
};\end{lstlisting}
	\caption{
		Contents of a \gls{VIRTIO} descriptor. The device-controlled value of \texttt{len} is not validated and is used as an argument to \texttt{memcpy}.}
\label{code:virtio_desc}
\end{figure}

When a \gls{VM} uses AMD \gls{SEV} or Intel \gls{TDX}, the \gls{VIRTIO} implementation relies on the DMA API and its swiotlb implementation.
The swiotlb layer allocates a shared memory region used for communication.
A device copies a payload to this region, the swiotlb layer performs some validations and finally \textit{bounces} the payload to the driver's private buffer when memory is synchronized.
The length of the copied data is given by the \texttt{len} field of the \texttt{vring\_desc} structure, shown in Figure~\ref{code:virtio_desc}.

Our fuzzing tool uncovered that the swiotlb layer does not validate the device-controlled length of the memory copy, which causes a buffer overflow.
This allows a malicious device to overwrite the private memory of a protected \gls{VM} with arbitrary data of arbitrary length.
This vulnerability occurs in shared code by all Linux \gls{VIRTIO} drivers.
What area of private memory can be overwritten depends on the memory allocator used by the driver.
For example, \texttt{virtio\_console} and \texttt{virtio\_rng} would allow overflowing a buffer allocated from the Linux SLAB allocator.
We verified empirically that \texttt{net\_device} structures, e.g. allocated during the initialization of the \texttt{e1000} device,
are often allocated in proximity to the descriptor buffers allocated in \texttt{virtio\_console}.
In such cases the \gls{HV} can exploit the out-of-bounds bug to overwrite the contents of the structure and craft an exploit similar to the previous example.

To increase the reliability of the exploit, the \gls{HV} infers the host virtual addresses of the descriptor buffers as well as the \texttt{net\_device} structures
by observing changes in encrypted memory.
Based on the inferred addresses, the \gls{HV} selects a descriptor that points to a buffer that is allocated in proximity to one of the \texttt{net\_device} structures
and adjusts the descriptors \texttt{len} field to overwrite the \texttt{net\_device} memory area.
If no suitable locations are found the \gls{HV} re-attaches the devices until a suitable combination of descriptor buffer and \texttt{net\_device} addresses is detected.

\subsection{Performance Analysis}

In this section, we evaluate the performance of \coolname{}.
We compare \coolname{} against a state-of-the-art \gls{VM}-based approach.
Further, we examine whether \textit{delay reduction} and \textit{targeted interrupt injection} are effective in improving fuzzing performance.
We conducted all of our experiments on a machine equipped with an 8-Core Intel i7-7700 and 64GB of memory.
The analysis of each driver was restricted to one CPU core, to eliminate differences in performance due to the multithreaded design of \coolname{}.
In order to reduce randomness, we used an empty initial corpus and a fixed random seed for all evaluations.
The size of individual corpus files was limited to $10$kB and a timeout of 120 seconds per fuzzing iteration was used.
The harness modules used for the evaluation are listed in our public repository~\cite{fuzzsarepo}.

\begin{table}
  \caption{Comparison of executions per second and percentage of driver code covered,
    between Agamotto~\cite{agamotto} and \coolname{} without (-D) and with (-ND) delay reduction,
    averaged over three hours.
    The evaluation was performed using the basic harness module, which performs one driver load/unload cycle.
    The coverage is calculated as the percentage of covered \texttt{trace-pc} instrumented instructions divided by the total number of \texttt{trace-pc} instrumented instructions.
    }
  \begin{center}
    \resizebox{\columnwidth}{!}{
    \begin{threeparttable}
      \begin{tabular}{l r r r @{}c r r}
	\toprule
	& \multicolumn{3}{c}{{\# Executions / s}}
	&
	& \multicolumn{2}{c}{{\% Code Covered}}
	\\
	\cmidrule{2-4}
	\cmidrule{6-7}
	& \thead{Agamotto} & \thead{\coolname-D} & \thead{\coolname-ND}
	&
	& \thead{Agamotto} & \thead{\coolname-ND}
	\\

	\midrule
	          8139cp & 1.39  &  1336.00 &  1275.50 & &   4 &  11 \\
                    e100 & 0.45  &    64.00 &   710.50 & &   3 &  10 \\
                   e1000 & 0.73  &     1.00 &    66.50 & &   4 &  16 \\
                  e1000e & 0.74  &    28.00 &   886.00 & &   2 &   9 \\
                     gve & 1.06  &     1.00 &   732.19 & &   5 &  18 \\
                ne2k-pci & 0.35  &  6604.00 &  6332.00 & &   1 &   1 \\
                    nvme & 0.79  &     0.02 &     1.55 & &  11 &  25 \\
                  rocker & 0.64  &  2752.00 &  4227.50 & &   1 &   2 \\
                  sungem & 0.98  &  3144.00 &  6199.50 & &   6 &   8 \\
                  sunhme & 1.14  &   786.00 &  1250.50 & &  14 &  18 \\
                 vmxnet3 & 0.74  &  2557.00 &  2735.00 & &   7 &  13 \\
	\bottomrule
      \end{tabular}
    \end{threeparttable}
  }
  \end{center}
\label{tbl:agamotto}
\end{table}

To evaluate the performance improvement compared to previous \gls{VM}-based approaches, we compare \coolname{}
against Agamotto~\cite{agamotto}, which is a state-of-the-art \gls{PCI} device fuzzer.
Agamotto is similar to USBFuzz~\cite{Peng2020} in function and performance.
It utilizes a virtual device which generates a mutated IO-stream that is
forwarded to the target driver. In addition, Agamotto uses dynamic checkpoints to speed up the analysis.
The evaluation was performed using a simple harness that loads and unloads a target driver, which is the only harness supported by Agamotto.
Targeted interrupt injection was enabled for each driver.
Each driver was analyzed for three hours and no modifications were applied to the driver code.
The results of the evaluation can be seen in Table~\ref{tbl:agamotto}.
On average \coolname{} improves the number of executed units per second by a factor of $1915$ and $2706$ for the delayed and non-delayed versions respectively.
In addition,
we calculate the percentage of driver code covered by dividing the number of covered \texttt{trace-pc} instrumented instructions by the
number of all \texttt{trace-pc} instrumented instructions within the respective driver.
On average \coolname{} improves the percentage of covered driver code by a factor of $2.26$.
During the analysis \coolname{} repeatedly triggered two bugs in the \texttt{gve} and \texttt{nvme} drivers, while Agamotto triggered none.

\begin{table}
    \caption{
      Comparison of executions per second and code block coverage as reported by \textit{libfuzzer} without (-D) and with (-ND) delay reduction,
      averaged over three hours.
      The evaluation was performed using the extended harness modules, which interact with resources instantiated by the target drivers.
    }
  \begin{center}
    \begin{threeparttable}
      \begin{tabular}{l r r@{ }r @{}c r r}
	\toprule
	& \multicolumn{3}{c}{{\# Executions / s}}
	&
	& \multicolumn{2}{c}{{\# Blocks}}
	\\
	\cmidrule{2-4}
	\cmidrule{6-7}
	& \thead{\coolname-D} & \multicolumn{2}{c}{\thead{\coolname-ND\\\bf{(Increase)}}}
	&
	& \thead{\coolname-D} & \thead{\coolname-ND}
	\\
	\midrule

	   8139cp &    1.32 &  122.41 &    $\times$\bf{92.58} & & 1038 &   1040 \\
	     acpi &    8.00 &   8.00  &           $\times$1.0 & &  71  &     71 \\
             e100 &   63.19 &  231.98 &     $\times$\bf{3.67} & &  573 &    569 \\
            e1000 &    3.00 &  259.06 &    $\times$\bf{86.35} & & 1427 &   1535 \\
           e1000e &    0.70 &  111.25 &   $\times$\bf{158.92} & & 1386 &   1579 \\
              gve &    2.00 &  636.22 &   $\times$\bf{318.11} & &  147 &    594 \\
         ne2k-pci & 1408.00 & 1658.00 &     $\times$\bf{1.18} & &   31 &     31 \\
             nvme &    0.02 &    0.88 &     $\times$\bf{44.0} & &  260 &    291 \\
      qemu-fw-cfg & 1254.00 &  1341.0 &     $\times$\bf{1.06} & &   35 &     37 \\
           rocker &  171.01 &  203.25 &     $\times$\bf{1.19} & &  181 &    184 \\
           sungem &    6.01 &   59.04 &     $\times$\bf{9.82} & &  924 &   1032 \\
           sunhme &  195.00 &  428.00 &     $\times$\bf{2.19} & & 1025 &   1030 \\
          tpm-tis &    2.00 &  857.00 &   $\times$\bf{428.50} & &  150 &    326 \\
      vio-balloon & 1291.00 & 1328.00 &     $\times$\bf{1.03} & &  281 &    281 \\
          vio-blk &  625.00 &  624.00 &          $\times$1.00 & &  333 &    333 \\
      vio-console &  349.00 &  444.00 &     $\times$\bf{1.27} & &  352 &    352 \\
       vio-crypto &  270.00 &  277.00 &     $\times$\bf{1.03} & &  258 &    258 \\
        vio-input &  393.00 &  635.00 &     $\times$\bf{1.62} & &  299 &    299 \\
          vio-net &  553.00 &  400.00 &          $\times$0.72 & & 1250 &   1257 \\
          vio-rng &    1.00 & 2282.00 &  $\times$\bf{2282.00} & &  238 &    239  \\
          vmxnet3 &   37.07 &   59.94 &     $\times$\bf{1.62} & &   51 &     51 \\

	\bottomrule
      \end{tabular}
    \end{threeparttable}%
  \end{center}
\label{tbl:dynamic-execs}
\end{table}

To evaluate the improvement through delay reduction, we measured the difference in code block coverage and executions per second with and without the optimization. %
We ran one instance of \coolname{} for each driver listed in Table~\ref{tbl:all_bugs}.
The execution was terminated after three hours, or after $128$ restarts in cases where bugs were triggered shortly after the start of the analysis.
To improve the analysis results, we patched identified bugs when possible, as detailed in our repository~\cite{fuzzsarepo}.
For the evaluation of network drivers, we used the more elaborate harness components for network, character and block device drivers, that are provided by \coolname{}.
These components load the driver, obtain the resources created by the driver, perform a small set of system calls on the resources and unload the driver again.
Each harness contains a concurrent thread that blocks until a workload initiated by the driver enters a blocked state,
in which case the routine is woken to trigger an interrupt.
All the \gls{VIRTIO} drivers were analyzed in \textit{simulation} mode.

Table~\ref{tbl:dynamic-execs} shows the difference in execution speed between the optimized and the un-optimized version.
To account for differences in performance due to increased code coverage, we additionally list the total number of covered code blocks.
The table shows that reducing delays issued by driver code dramatically increases performance for most drivers without a negative impact on code coverage.
We measure an average performance improvement by a factor of $163$ through reducing delays.
However, in some cases, we see either no significant improvement or even a slight reduction in performance.
We determined two distinct causes of this behavior.
In the first case, we see that the code block coverage for drivers with little or no improvement is low.
Therefore, it is likely that the driver initialization failed early before any major interaction with the device was performed.
The second case covers \gls{VIRTIO} drivers.
These drivers are specifically designed for software device back-ends and do not rely on timeouts as much as device drivers that can also interact with real hardware.
We recommend disabling delay optimization in cases where it does not increase overall performance.
\definecolor{dodgerblue4}{rgb}{0.061,0.231,0.471}
\begin{figure}
  \centering
\resizebox{\columnwidth}{!}{%
\begin{tikzpicture}
    \begin{axis}[
	ybar,
	bar shift=0.0cm,
        xlabel={Seconds},
        ylabel={Frequency},
        xmin=1.0,
	ymode=log,
	xmode=log,
        ymin=1,
	y=0.3cm,
	bar width=0.1cm,
      ]
\addplot[black, fill=dodgerblue4, fill opacity=1.0, draw=none] table [
   col sep=comma,
   y=tiq0-1-1000,
   x=Count,
]{./figures/data/irq.csv};
\addlegendentry{Random}
\addplot[black, fill=lightgray, fill opacity=0.75, draw=none] table [
   col sep=comma,
   y=tiq1-1-1000,
   x=Count,
]{./figures/data/irq.csv};
\addlegendentry{Targeted}

\end{axis}
\end{tikzpicture}}
  \caption{Distribution of the time in seconds required to trigger a bug in the \texttt{rocker} driver with targeted and random interrupt injection.}
  \label{fig:irq}
\end{figure}
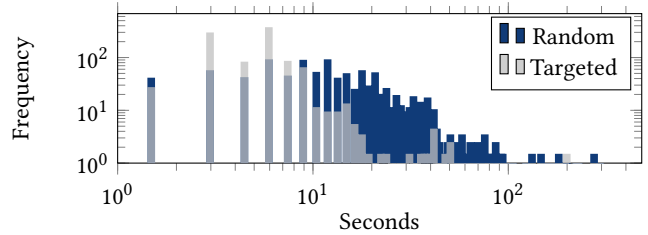

We focused on the \texttt{rocker} driver to evaluate whether targeted interrupt injection improves the bug-finding capabilities of \coolname.
The \texttt{rocker} driver was selected for this evaluation
because it requires multiple interrupts to be triggered during its initialization.
If no interrupt is triggered within a fixed time window, the initialization fails.
To evaluate the effectiveness of the optimization, we ran two instances of \coolname:
one instance used targeted interrupt injection; the other injected interrupts at random time intervals ranging from $1-1000$ nanoseconds.
The experiment was repeated over $1000$ runs and each run was started with an empty corpus.
Figure~\ref{fig:irq} compares the average time in seconds required to trigger the first bug.
Overall, the optimized version took on average of $6.78$ seconds to trigger the bug, whereas the unoptimized version took $17.66$ seconds.
Targeted interrupt injection thereby reduces the time required to trigger the bug by a factor of $2.6$.

To evaluate the overall effectiveness of \coolname{} in uncovering bugs in the \gls{HV} interface of device drivers,
we measured the elapsed time until the first bug is triggered and averaged the measurement over $128$ runs.
For this evaluation, we applied the modifications to remove complex checks on data provided by the fuzzer as detailed above.
In addition, we applied patches to avoid \textit{assertion failures}, \textit{unbounded allocations} and \textit{deadlocks}.
Table~\ref{tbl:ttb} shows the results from this evaluation.
For most drivers, the first bug is triggered within the first few minutes of fuzzing,
with many bugs being triggered in under one minute of analysis.
This result demonstrates the effectiveness of \coolname{} in finding such issues,
but also underlines the extent of missing sanitization of data provided by the \gls{HV} in the Linux kernel code-base.

\begin{table}
  \caption{%
  	Time required to trigger the first Bug (TTB) in seconds for each examined driver with a known bug,
	averaged over 128 runs.
	Patches were applied to avoid low severity bugs due to assertion failures, unbounded allocations and deadlocks.
      	}
  \begin{center}
\resizebox{\columnwidth}{!}{%
    \begin{tabular}{l r | l r | l r}
      \toprule
	    & \thead{TTB (s)}
      &
	    & \thead{TTB (s)}
      &
	    & \thead{TTB (s)} \\
      \midrule
          8139cp &                         4.60 &         acpi &                     1.87 &        e100 &             1424.80 \\
	   e1000 &                        51.54 &       e1000e &                  1650.98 &         gve &              542.82 \\
	    nvme &                         0.37 &      qemu-fw &                    2.03  &      rocker &                6.78 \\
	  sungem &                         1.50 &      tpm-tis & \textsuperscript{*}7200+ & vio-balloon &               37.86 \\
	 vio-blk &                        20.24 &  vio-console &                    18.28 &  vio-crypto &               18.92 \\
	 vio-net &                         4.93 &     vio-ring &                   132.34 &     vio-rng &                2.08 \\
	 vmxnet3 &                         1.89 &              &                          &             &                     \\

      \bottomrule
    \end{tabular}}
  \end{center}
  \begin{tablenotes}
  \item \scriptsize{\textsuperscript{*} Estimate based on few runs, since the bug could not be triggered reliably.}
  \end{tablenotes}
  \label{tbl:ttb}
\end{table}

\section{Related Work}

\paragraph{Hardware Interface Analysis}
Work on analyzing the hardware side interface of device drivers has mainly focused on the USB interface, with limited support for PCI and other hardware buses.
Generic approaches are usually based on symbolic execution or require specific hardware devices.

Syzkaller-USB~\cite{github2019syzkaller} adapts the Linux kernel to inject USB packets via the system call interface directly.
In order to avoid the requirement for such modifications to the target operating system, a virtual device can be used to forward the fuzzing input. %
This approach is implemented by USBFuzz~\cite{Peng2020} and Spenneberg et al.~\cite{Spenneberg2014DonT} to analyze USB device drivers.
Agamotto~\cite{agamotto} implements a similar approach using a virtual PCI device. %
Keil et al.~\cite{Keil_statefulfuzzing} use a virtual device to emulate a specific WiFi chip in order to analyze its respective device driver.
Reviewing these approaches, we found their performance to be insufficient.
Both Agamotto and USBFuzz cite an average of multiple seconds of execution time per test case.
Further, none of the existing techniques can be applied to directly target the \gls{DMA} interface and lack
support for relevant device classes such as \gls{VIRTIO} and Platform.

Symdrive~\cite{symdrive}, POTUS~\cite{206162} and DDT~\cite{Kuznetsov} facilitate the symbolic execution capabilities provided by S2E~\cite{s2e}
to systematically explore driver state.
While symbolic execution can analyze hardware interfaces in a generic way,
performance is often limited due to path explosion and expensive constraint solving.
SADA~\cite{272210} proposes a static approach to detect several classes of unsafe \gls{DMA} accesses combined with lightweight symbolic execution.
While SADA improves the performance issues of pure symbolic execution methods it incurs disadvantages due to the static nature of the analysis.
For example, bugs depending on the runtime state of the driver such as bugs caused by improper handling of failed driver initialization are difficult to detect with purely static approaches.
Further, due to incomplete symbolic analysis SADA produces false positives which have to be manually analyzed.

Periscope~\cite{Song2019} presents a bus agnostic approach to intercept \textit{on device} communication between hardware and an operating system and uses this method to implement a fuzzer.
To analyze a driver Periscope requires the corresponding device and only analyzes bottom-half interrupt handlers.
Driver (de-)initialization, interrupt generation and top-half interrupt handling are excluded from the analysis.
Markettos et al.\cite{Markettos2019} describe a custom hardware component that can be attached to the PCI port in order to trigger bugs in device drivers communicating over that interface.
A similar approach based on custom hardware is presented by TTWE~\cite{185177}.
In comparison to these approaches, \coolname{} does not require specific hardware.

\paragraph{Virtualization-Free Kernel Fuzzing}
Janus~\cite{Xu2019FuzzingFS} and Hydra~\cite{Kim2019FindingSB} utilize the userspace kernel environment provided by \gls{LKL} to quickly reset state during fuzzing via \texttt{fork}.
Based on this primitive, they build a dynamic analysis framework targeted at finding bugs in Linux file system implementations.
These approaches, however, target file system implementations and lack support for hardware interface analysis.

\paragraph{Attacks on SEV and TDX}
Researchers have already presented security issues in AMD \gls{SEV} and \gls{SEV-ES} under the threat model of a malicious \gls{HV}.
The discovered vulnerabilities rely on the missing protection of Second Level Address Translation~\cite{hetzelt2017security, morbitzer2018severed} and missing memory integrity protection from software attacks~\cite{du2017secure, li2019exploiting, wilke2020sevurity}.
Both issues are addressed in the designs of AMD \gls{SEV-SNP} and Intel \gls{TDX}.
In comparison, the vulnerabilities presented in Section~\ref{sec:case-study} are purely software bugs and cannot be protected against through only memory encryption and integrity protection.
The discovered bugs do not apply to %
a particular confidential computing solution, but cover all solutions which rely on devices provided by a malicious \gls{HV}.
Radev and Morbitzer~\cite{radev2020exploiting} discovered new attack vectors for \gls{SEV-ES} \glspl{VM} based on trusting the \gls{HV} with the emulation of few special instructions, the creation of \gls{MMIO} regions, and the attachment of VirtIO devices.
While the authors suggest that trusting virtualized devices can impact the security of a \gls{SEV-ES} \gls{VM}, they only examine the risks for the \gls{VM} in delegating cryptographic operations to the device \texttt{virtio-crypto} and in using the device \texttt{virtio-rng} as a source of entropy.
Our work also confirms the security implications for protected \glspl{VM} introduced by devices attached to the \gls{VM}.
However, we also provide an automated, scalable approach for discovering software bugs in Linux drivers. %

\section{Discussion}

\paragraph{Bug Exploitability}
Software bugs in a protected \gls{VM} carry a high risk since a malicious \gls{HV} holds fine-grained control over the \gls{VM}'s execution from the initial boot stage~\cite{radev2020exploiting}.
For example, a malicious \gls{HV} may track the \gls{VM}'s execution to execute an attack at the precise time~\cite{radev2020exploiting}, infer address space information via side-channels~\cite{li2019exploiting}, or employ multiple malicious virtual devices to combine bugs in various drivers.
The exploitability of the discovered bugs and exploitation steps are agnostic to the \gls{VM} protection technology, but rather depend on the \gls{VM}'s Linux kernel self-protection configuration~\cite{linux-self-protection}.
For example, all discovered \textit{device-shared pointer} bugs are exploitable if the virtual offset of the Linux kernel image becomes known to the attacker via another bug or side-channel.

Reducing the trusted computing base can be an effective approach for reducing the amount of software which contains exploitable bugs and requires validations under this new threat model.
The early publicly available implementation of Intel \gls{TDX} attempts to limit the number of buggy device drivers by adding a \textit{device filter}~\cite{github2020tdxdevfilter}.
The filter blocks discovery of \gls{PCI} and ACPI devices that are not present in a pre-compiled list.
However, the discovered \textit{VirtIO / swiotlb} bugs from Section~\ref{sec:case-study} are applicable to the VirtIO devices featured in the Intel \gls{TDX} device filter at the time of writing.
Additionally, cloud providers may have different device and software stack requirements, which would require adjustments to the list of allowed devices.
For example, we determined that the Google Confidential \gls{VM} does not utilize any of the allowed devices in the \gls{TDX} device filter.

\paragraph{State Accumulation}
\coolname{} avoids state accumulation by re-loading the analyzed driver after each fuzzing iteration.
However, some state might still persist and influence future iterations.
In cases where state accumulation becomes a problem the fuzzing process can be forked as described by Xu et al.~\cite{Xu2019FuzzingFS} to provide a clean state for each new iteration.
\coolname{} supports libfuzzer's \textit{Fork Mode}~\cite{libfuzzer} to periodically reset the state of the kernel environment.

\paragraph{Limitations}
\gls{LKL} handles concurrency by managing different posix threads that are synchronized via semaphores.
If one thread terminates or stalls its execution, it releases the semaphore, and a new thread can be scheduled.
Therefore, there is no implicit preemption in \gls{LKL}, which limits its capabilities in finding race conditions.

As described in Section~\ref{sec:eval}, \coolname{} requires complex checks such as checksum comparisons to be removed manually.
While we do not address such issues directly in this work, we believe that by moving the analysis to
userspace, \coolname{} can benefit more readily from recent advances in the field.
For example, approaches based on dataflow tracing~\cite{Ryan2019} are currently only available in userspace and can be integrated into \coolname{}.

\paragraph{Disclosure Response}
The discovered bugs were reported to several entities including AMD, Intel, Google, driver maintainers and the Linux kernel security, network and virtualization mailing lists.
It was recommended to publicize discovered issues, since the threat model of current Linux releases does not yet consider malicious virtual devices.
As the bugs do not affect core products of producers and customers of protected virtual machine technologies, they need to be addressed by driver maintainers.
Several of the bugs uncovered in \texttt{swiotlb}, \texttt{virtio\_*}, \texttt{vmxnet3} and \texttt{gve} drivers have been addressed by the respective maintainers.

\section{Conclusions}

We presented \coolname{}: a framework for dynamic analysis of drivers under the threat model of a malicious \gls{HV}.
\coolname{} was applied to \drvanalyzed drivers in Linux \kernelversion, which covered network drivers, \gls{VIRTIO}-based drivers, and platform drivers.
The evaluation found \nbugs bugs from a variety of classes and initiated multiple patches to the Linux kernel.
We implement Proof-of-Concept exploits for three bug classes that gain code execution or corrupt memory inside an AMD \gls{SEV-ES} \gls{VM}.
The described exploits demonstrate how the capabilities of a malicious \gls{HV}, such as control over virtual devices, fine grained control over \gls{VM} execution as well as information about the \gls{VM} execution state,
can be used to craft reliable exploits.
The majority of bugs were reported to the driver maintainers, and the security implications were responsibly disclosed.

Unlike other solutions for driver analysis, \coolname{} carefully considers the capabilities of the \gls{HV} and the semantics of the \gls{DMA} API to detect more cases of improper input sanitization and anti-patterns in device programming.
\coolname{} provides a coverage-driven in-process userspace fuzzer built upon libfuzzer and \gls{LKL}.
Compared to existing approaches, \coolname{} reduces setup overhead by moving the analysis into a userspace program.
In combination with targeted optimizations of the userspace kernel environment, \coolname{} drastically improves dynamic analysis throughput.
To the best of our knowledge, \coolname{} presents the first approach to analyze the device driver interface under the new threat model of protected virtual machines.

Our results suggest that many Linux device drivers do not correctly sanitize device-provided data.
Under the threat model of AMD \gls{SEV-SNP} and Intel \gls{TDX}, this presents a serious security risk for the protected \gls{VM}, which may be exploited through a software driver bug.
Thus, the software operating under such a threat model should reduce the interfaces with untrusted entities and should rely on established methods for testing the remaining interfaces.

\section{Acknowledgments}
The authors would like to thank our shepherd, Fengwei Zhang, and the anonymous reviewers for their valuable feedback.
This material is based upon work partially supported by the European Commission under the Horizon 2020 Programme (H2020) as part of the LOCARD project (G.A. no. 832735).
Any opinions, findings, and conclusions or recommendations expressed in this material are those of the authors and do not necessarily reflect the views of our funding agencies.

\bibliographystyle{ACM-Reference-Format}
\bibliography{\jobname}
\end{document}